\documentclass[superscriptaddress,orcidlink,
 reprint,
 nofootinbib,
 amsmath,amssymb,
 aps,
]{revtex4-1}
\usepackage{orcidlink}
\usepackage{graphicx}
\usepackage{dcolumn}
\usepackage{bm}
\usepackage{amsmath}
\usepackage{enumitem}
\usepackage{subcaption}
\usepackage{float}
\usepackage{ulem}
\usepackage{verbatim}

\begin{document}
\title{Impact of LRG1 and LRG2 in DESI 2024 BAO data on dark energy evolution}

\author{Guanlin Liu \orcidlink{0009-0000-9225-7169}}
\email{irithyll0110@mail.ustc.edu.cn}
\affiliation{CAS Key Laboratory for Researches in Galaxies and Cosmology, Department of Astronomy, University of Science and Technology of China, Chinese Academy of Sciences, Hefei, Anhui 230026, China}
\affiliation{School of Astronomy and Space Sciences, University of Science and Technology of China, Hefei 230026, China}

\author{Yu Wang}
\affiliation{CAS Key Laboratory for Researches in Galaxies and Cosmology, Department of Astronomy, University of Science and Technology of China, Chinese Academy of Sciences, Hefei, Anhui 230026, China}
\affiliation{School of Astronomy and Space Sciences, University of Science and Technology of China, Hefei 230026, China}

\author{Wen Zhao \orcidlink{0000-0002-1330-2329}}
\email{wzhao7@ustc.edu.cn}
\affiliation{CAS Key Laboratory for Researches in Galaxies and Cosmology, Department of Astronomy, University of Science and Technology of China, Chinese Academy of Sciences, Hefei, Anhui 230026, China}
\affiliation{School of Astronomy and Space Sciences, University of Science and Technology of China, Hefei 230026, China}

\begin{abstract}
Recent measurements of baryon acoustic oscillations (BAO) by the Dark Energy Spectroscopic Instrument (DESI) suggest a preference for a dynamic dark energy model over a cosmological constant. This conclusion emerges from the combination of DESI's BAO data with observations of the Cosmic Microwave Background (CMB) and various type Ia supernova (SN Ia) catalogues. The deviation observed in the cosmological constant ($\Lambda$) reflects a departure from the standard cosmological model. Testing this deviation involves examining the consistency between cosmological parameters derived from early and late-time observations. Specifically, we focus on the matter density parameter $\omega_m = \Omega_mh^2$ and introduce ${\rm ratio}(\omega_m)$ to assess consistency, which is defined as the ratio of $\omega_m$ values constrained by high and low-redshift measurements. This ratio serves as a metric for quantifying deviations from the $\Lambda$CDM model. In this paper, we find that the DESI BAO+CMB yields ${\rm ratio}(\omega_m)=1.0171\pm0.0066$. Upon excluding the LRG1 and LRG2 data in DESI BAO, this ratio adjusts to ${\rm ratio}(\omega_m)=1.0100\pm0.0082$. This shift, corresponding to a change from $2.6\sigma$ to $1.2\sigma$, indicates that the deviation from the $\Lambda$CDM model is predominantly driven by these two samples from the DESI BAO measurements. To substantiate this conclusion, we utilized two cosmological model-independent methods to reconstruct the cosmic expansion history. Both reconstructions of the Hubble parameter $H(z)$ indicate that the evolving features of dark energy are determined by the combined LRG1 and LRG2 data. Therefore, different methods have reached the same conclusion, namely the importance of accurately measuring the BAO feature in LRG1 and LRG2 data. 

\end{abstract}

\maketitle

\section{Introduction} \label{sec1} 

The 6-parameter standard cosmological model, $\Lambda$CDM, assumes radiation and matter, plus cold dark matter and a cosmological constant $\Lambda$ in an isotropic and homogeneous universe. It provides an excellent agreement between theory and measurements of the power spectra of the temperature and polarization of the thermal cosmic background (CMB) radiation \cite{2020planck, aiola2020act, balkenholspt}, as well as other observations such as Type Ia supernovae (SN Ia) \cite{scolnic2018pantheon, brout2022pantheon+} and large-scale structure \cite{alam2017sdss3, alam2021completed, abbott2018desy1, kazin2014wigglez}. However, recent phenomena like the Hubble tension and the $S_8$ tension \cite{verde2019tensionof5sig, abdalla2022cosmologyreview,Blanchard_2024addition3} have challenged the $\Lambda$CDM model. Additionally, the key features of the model, such as the nature of dark energy driving the accelerated expansion of the universe, are poorly understood at a fundamental level. These challenges drive us to search for new physics beyond $\Lambda$CDM through more precise observations.

The Dark Energy Spectroscopic Instrument (DESI) \cite{aghamousa2016desi1, abareshi2022overview_bao} is conducting a Stage IV survey \cite{albrecht2006DETF} aimed at significantly enhancing cosmological constraints through measurements of the baryonic acoustic oscillation (BAO) feature embedded in the clustering of matter. Recently, the DESI collaboration released Data Release 1 (DR1) of BAO measurements from galaxy, quasar, and Lyman-$\alpha$ forest tracers \cite{adame2024desi2, adame2024desi3}. The corresponding cosmological analysis of these BAO datasets is detailed in Ref.\cite{adame2024desibao}, where no evidence for a dark energy equation of state (EoS) $w \neq -1$ was found under the assumption of a flat wCDM model. However, within the $w_0w_a$ cosmology framework \cite{chevallier2001CPL1, linder2003CPL2}, defined as $w(a) = w_0 + w_a(1-a)$ where $a$ is the scale factor, the inclusion of supernova samples with Planck CMB \cite{2020planck} and DESI BAO data revealed a preference for evolving dark energy compared to $\Lambda$CDM at approximately the $2.5\sigma$ (Pantheon$+$ \cite{brout2022pantheon+}), $3.5\sigma$ (Union 3 \cite{rubin2023union3}), and $3.9\sigma$ (DESY5 SN \cite{abbott2024DESY5}) levels, which is more significant than the constraints obtained from non-DESI BAO data \cite{park2024nonDESIdataw0wa}.

There are numerous works explaining the physical reasons for the time evolution of dark energy \cite{lodha2024desi_physics, yin2024desi_physics2, colgain2024desi_physics3, carloni2024desi_physics4,yang2024desi_physics5}, as well as discussions on the feasibility of quintessence dark energy models \cite{shlivko2024quintessence1, berghaus2024quintessence2, tada2024quintessence3,gialamas2024addition1}. However, there are $1.6\sigma$ and $2.2\sigma$ inconsistencies of unknown origin \cite{adame2024desibao} between the DESI LRG1 ($z_{\rm eff}=0.51$) and LRG2 ($z_{\rm eff}=0.71$) data and the BOSS DR12 results at similar redshifts. Given that these redshifts fall within the dark energy-dominated epoch, the combination of DESI LRG1 and LRG2 data may contribute to the dynamical features of dark energy. This reminds us that the indications of new physics in dark energy could merely be statistical fluctuations \cite{efstathiou2024challenges}. Therefore, it is necessary to determine the influence of the combination of DESI LRG1 and LRG2 data on the evolution of dark energy. In this paper, we use the parameter ${\rm ratio}(\omega_m)$ developed in our previous work \cite{Liu2024mine}, which can detect departures from the standard $\Lambda$CDM cosmological model by testing the consistency of early and late-time parameters. Specifically, ${\rm ratio}(\omega_m)$ is the ratio of the parameter $\omega_m$ constrained by high-redshift and low-redshift observations, which theoretically equals 1 under $\Lambda$CDM. We find ${\rm ratio}(\omega_m) = 1.0171 \pm 0.0066$ for DESI and ${\rm ratio}(\omega_m) = 1.0100 \pm 0.0082$ for DESI excluding LRG1 and LRG2, showing a reduction from $2.6\sigma$ to $1.2\sigma$, supporting that the combination of DESI LRG1 and LRG2 data leads to the dynamical features of dark energy.

The authors of Ref.\cite{wang2024self, wang2024role,dinda2024addition2} found that the LRG1 sample hardly affects the constraints on dark energy, while the LRG2 sample predominantly favors the presence of dynamical dark energy through a step-by-step exclusion of data points, assuming a flat $w_0w_a$CDM model. However, we find ${\rm ratio}(\omega_m) = 1.0124 \pm 0.0071$ for DESI data excluding LRG1, which still shows a $1.7\sigma$ deviation from $\Lambda$CDM. This discrepancy suggests that the behavior of dark energy evolution may be influenced by model dependence. Therefore, we employed two model-independent methods, the Taylor reconstruction (TR) of $H(z)$ developed in Ref.\cite{gu2024dynamical, wang2024dynamical} and the Chebyshev reconstruction (CR) of $w(z)$ developed in Ref.\cite{calderon2024desi}, to extract information on dark energy evolution from the observational data. Through the reconstruction results of different data combinations, we found that the BAO observations at the redshift positions of DESI LRG1 and LRG2 are crucial for determining the evolution of dark energy and the late-time expansion history of the universe.

This article is organized as follows. In Section \ref{sec2}, we introduce the methods and datasets used in this analysis. The results and related discussions are presented in Section \ref{sec3}. Finally, Section \ref{sec4} presents our conclusions in this article.

\section{Methodology and Data} \label{sec2}

In this section, we will introduce the methods and data used in this analysis.

\subsection{${\rm ratio}(\omega_m)$ for Testing $\Lambda$CDM}
In the standard cosmological model, the expansion history of the Universe is governed by the Friedmann equation
\begin{eqnarray}\label{LCDM_H}
    E(z)&\equiv& H(z)/H_0 \nonumber \\
    &=&\sqrt{{\Omega}_r(1+z)^4+{\Omega}_m(1+z)^3+{\Omega}_k(1+z)^2+{\Omega}_{\Lambda}} ,\nonumber
\end{eqnarray}
where ${\Omega}_r$, ${\Omega}_m$, ${\Omega}_k$ and ${\Omega}_{\Lambda}$ are the fractional densities of radiation, matter, spatial curvature and dark energy at redshift $ z = 0 $ and constrained by $ {\Omega}_r+{\Omega}_m+{\Omega}_k+{\Omega}_{\Lambda}=1 $. The Hubble parameter, $H(z)$, characterizes the expansion rate of the Universe at a redshift $z$, while $H_0$ represents the current expansion rate, commonly referred to as the Hubble constant. The function $E(z)$ encapsulates the contributions and evolution of different components within the Universe. In this study, we assume a flat universe, which correspond to ${\Omega}_k = 0$. 

As discussed above, the cosmological parameters $\Omega_m$ and $H_0$ are independent of time. Therefore, their values constrained by high-redshift observations should be consistent with those derived from low-redshift observations. Note that the recombination redshift $z_*$ serves as the dividing line between the early and late Universe. Parameters denoted with a tilde (e.g., $\tilde{\Omega}_m$, $\tilde{H}_0$) are inferred from observations at redshifts greater than the recombination redshift $z_*$ and are referred to as early cosmological parameters. Conversely, parameters without a tilde (e.g., ${\Omega}_m$, $H_0$) are considered as late cosmological parameters.

In previous work \cite{Liu2024mine}, we selected the cosmological parameter $\omega_m = \Omega_mh^2$ as a representative to analyze the potential difference between the early and late universe due to parameter degeneracy. Here, $H_0=100h\ \rm km\ s^{-1}\ Mpc^{-1}$ represents the dimensionless Hubble constant. Within the $\Lambda$CDM framework, ${\rm ratio}(\omega_m) \equiv \tilde{\omega}_m / \omega_m$ is expected to equal $1$. Consequently, the value of ${\rm ratio}(\omega_m)$ serves as an indicator of any deviations from the $\Lambda$CDM model. In practice, the combination of DESI 2024 BAO data and \textit{Planck} CMB data provides a stringent constraint on ${\rm ratio}(\omega_m)$. The details of this constraint will be discussed in Section \ref{sec3}.

\subsection{Reconstruction of the Hubble Parameter}
The Hubble parameter $H(z)$ is the expansion rate of the Universe at redshift $z$, which describes the evolution of the energy density of different components. Therefore, employing cosmological model-independent methods to reconstruct $H(z)$ from observational data can aid in extracting robust and valuable information from the history of cosmic evolution. Given our objective to investigate the characteristics of evolving dark energy, we utilize the Taylor reconstruction method as developed in Ref.\cite{gu2024dynamical, wang2024dynamical}, along with the Chebyshev reconstruction method \footnote{Actually, this method is independent of any specific dark energy model, but it remains within the framework of the $\Lambda$CDM paradigm.} detailed in Ref.\cite{calderon2024desi}. In the following sections, we will provide an introduction to these two model-independent reconstruction techniques.

\subsubsection{Taylor Reconstruction} 
Following Ref.\cite{zhu2015optimal}, the distance–redshift relation can be expressed as a small fluctuation about a fiducial model and expanded as a Taylor series around $x=0$
\begin{eqnarray}\label{TR_chi}
    \frac{\chi(z)}{\chi_{\rm fid}(z)}&=&\alpha_0\left[1+\alpha_1 x+\frac{1}{2}\alpha_2 x^2+\frac{1}{6}\alpha_3 x^3+\frac{1}{24}\alpha_4 x^4 \right], \nonumber \\ 
    x &\equiv& \frac{\chi_{\rm fid}(z)}{\chi_{\rm fid}(z_p)}-1,
\end{eqnarray}
where $z_p$ is the pivot point for the expansion and $\chi(z)$ is the comoving distance
\begin{equation}\label{chi_normal}
    \chi(z)= \int_{0}^{z}\frac{c}{H(z')}\mathrm{d}z'.
\end{equation}
Here $\chi_{\rm fid}(z)$ is the comoving distance derived from a fiducial model, which is a flat $\Lambda$CDM with parameters constrained by the \textit{Planck} measurements \cite{2020planck}. As indicated in Ref.\cite{zhu2015optimal,gu2024dynamical,wang2024dynamical}, the total analysis does not depend on a particular choice of $z_p$ and we adopt the same pivot point $z_p=0.5$ as used in Ref.\cite{wang2024dynamical}.
It is essential to note that the selection of the expansion terms plays a crucial role in balancing the bias and uncertainty in the reconstruction process. In our analysis, we follow the choice of Ref.\cite{wang2024dynamical} and expand the series up to the fourth order. Subsequently, the five parameters $\alpha_0$ through $\alpha_4$, along with the variable $x$ defined in Eq.\eqref{TR_chi}, are utilized to express $H(z)$
\begin{eqnarray} \label{TR_H}
    \frac{H_{\rm fid}(z)}{H(z)} = \alpha_0 [ 1&+&\alpha_1+(2\alpha_1+\alpha_2)x+\frac{1}{2}(3\alpha_2+\alpha_3)x^2 \nonumber \\ 
    &+&\frac{1}{6}(4\alpha_3+\alpha_4)x^3+\frac{5}{24}\alpha_4x^4 ].
\end{eqnarray}
As a result, once the five Taylor coefficients which can express $H(z)$ and $\chi(z)$ are constrained by the distance measurements, $H(z)$ can be reconstructed in the low-redshift range independent of any specific cosmological model.

\subsubsection{Chebyshev Reconstruction} 
In the appropriate Sobolev space, the set of Chebyshev polynomials form an orthonormal basis. Thus, any function within this space can be expanded as a series over the interval $-1 \leq x \leq 1$ as follows
\begin{equation} \label{Chebyshev expansion}
    y(z)= \sum_{i=0}^N C_i T_i(x),\qquad x\equiv 1-2\frac{z_{\rm max}-z}{z_{\rm max}-z_{\rm min}}\in[-1,1]
\end{equation}
where $T_i(x)$ are the Chebyshev polynomials of the first kind, defined in the interval $x\in[-1, 1]$, and $C_i$ are the coefficients associated with these polynomials. In the limit as $N \to \infty$, the Chebyshev series forms a complete basis for the space of all continuous functions. For practical purposes, we adopt the same redshift range as used in Ref.\cite{calderon2024desi}, specifically $z_{\rm min} = 0$ to $z_{\rm max} = 3.5$, to ensure that all observational data can contribute to determining the evolution in this redshift range. The authors of Ref.\cite{calderon2024desi} limit the analysis to four terms, and we follow this choice to balance reconstruction accuracy and precision. The first four Chebyshev polynomials are given by
\begin{eqnarray}
    T_0(x)&=&1,\qquad T_1(x)=x, \nonumber \\
    T_2(x)&=&2x^2-1,\qquad T_3(x)=4x^3-3x .
    \label{Chebyshev4}
\end{eqnarray}
Then, the equation of state of dark energy, $w(z)$, can be expressed as a deviation from $w = -1$.
\begin{equation}
    w(z)=-1 \times \sum_{i=0}^{N=3} C_i T_i(x).
    \label{w_Cheby}
\end{equation}
Once the four free parameters $C_0$ to $C_3$ are constrained by observational data, we can reconstruct $w(z)$, from which the effective dark energy densities $f_{\rm DE}(z)$ can be derived
\begin{eqnarray}\label{effective DE density}
    f_{\rm DE}(z)&\equiv& \frac{\rho_{\rm DE}}{\rho_{\rm DE,0}} \nonumber \\
    &=&  \exp\left[3\int_0^z (1+w(z'))\mathrm{d} \ln(1+z') \right].
\end{eqnarray}
Then, the reconstruction of $E(z)$ is also completed
\begin{equation}\label{CR_H}
    E(z)=\sqrt{{\Omega}_r(1+z)^4+{\Omega}_m(1+z)^3+{\rm \Omega}_{\Lambda}f_{\rm DE}(z)}.
\end{equation}

\subsection{Data} 
\subsubsection{BAO}
BAO are remnants of sound waves from the early universe, imprinted as a characteristic scale known as the sound horizon $r_s$ after recombination. This scale serves as a cosmic standard ruler for measuring the large-scale structure of the universe. 

The comoving sound horizon $r_s(z)$ is given by \cite{eisenstein1998baorigin}:
\begin{eqnarray}\label{rs_formula}
    r_s(z)=\frac{c}{H_0}\int_{z}^{\infty}\frac{c_s}{E(z')}\mathrm{d}z' ,
\end{eqnarray}
where $c$ is the speed of light, the sound speed is $c_s=1/\sqrt{3(1+R_ba)}$, with $R_ba=3\rho_b/(4\rho_r)$, and $R_b=31500\omega_b(T_{CMB}/2.7K)^{-4}$ with $\omega_b=\Omega_bh^2$ and the CMB temperature $T_{CMB} = 2.7255K$. It is crucial to account for the radiation when calculating the comoving sound horizon $r_s$. This component can be quantified by the matter fraction $\omega_m$ using the matter-radiation equality relation, $\Omega_r = \Omega_m / (1 + z_{eq})$, where $z_{eq}$ is given by $z_{eq} = 2.5 \times 10^4 \omega_m (T_{CMB} / 2.7,\text{K})^{-4}$. It is important to note that the computation of $r_s$ is based on pre-recombination information, implying that $r_s$ is only calculatable in our analysis when including CMB data.

\begin{table}[htb]
\captionsetup{justification=raggedright}
\begin{ruledtabular}
\begin{tabular}{ccccc}
 $z_{eff}$ & $D_M/r_d$ & $D_H/r_d$ & $r$ or $D_V/r_d$ & Tracer \\ \hline
    0.30 & - & - & $7.93\pm0.15$ & BGS \\       
    0.51 & $13.62\pm0.25$ & $20.98\pm0.61$ & $-0.445$ & LRG \\ 
    0.71 & $16.85\pm0.32$ & $20.08\pm0.60$ & $-0.420$ & LRG \\ 
    0.93 & $21.71\pm0.28$ & $17.88\pm0.35$ & $-0.389$ & LRG+ELG \\ 
    1.32 & $27.79\pm0.69$ & $13.82\pm0.42$ & $-0.444$ & ELG \\
    1.49 & - & - & $26.07\pm0.67$ & QSO \\ 
    2.33 & $39.71\pm0.94$ & $8.52\pm0.17$ & $-0.477$ & Ly$\alpha$ \\
\end{tabular}
\caption{\label{tab:DESI}A list of DESI BAO datasets used in this cosmological analysis is provided below. Here, $r$ represents the correlation coefficient between $D_M/r_d$ and $D_H/r_d$.}
\end{ruledtabular}
\end{table}

In Appendix E of Ref.\cite{hu1996analyticalformula}, fitting functions are developed to provide convenient and percent-level accurate approximations across a broad parameter space. These fitting functions for the drag epoch redshift $z_d$ and recombination redshift $z_*$ have been extensively adopted in subsequent studies \cite{komatsu2009five, wang2007distanceprior-1, wang2013distanceprior-2, wang2019distanceprior-3}. In our analysis, we also apply these fitting functions to calculate the sound horizon. The redshift of the drag epoch can be approximated as \cite{hu1996analyticalformula}
\begin{equation}\label{z_d}
    z_d=\frac{1345\omega_m^{0.251}}{1+0.659\omega_m^{0.828}}[1+b_1\omega_b^{b_2}],
\end{equation}
where
\begin{equation}\label{z_d_params}
    b_1=0.313\omega_m^{-0.419}[1+0.607\omega_m^{0.674}], b_2=0.238\omega_m^{0.223}.
\end{equation}

\begin{table*}[htb]
\captionsetup{justification=raggedright}
\begin{ruledtabular}
\begin{tabular}{ccccc}
 $z_{eff}$ & Measured Quantity & Value & Dataset & Reference \\ \hline
    0.122 & $D_V/r_d$  & $3.654\pm0.115$ & 6dFGS+SDSS MGS & \cite{beutler20116df,ross2015mgs} \\
    0.38 & $D_H/r_d$ & $25.00\pm0.76$ & SDSS DR12 & \cite{alam2021completed} \\
    0.38 & $D_M/r_d$ & $10.23\pm0.17$ & SDSS DR12 & \cite{alam2021completed} \\
    0.51 & $D_H/r_d$ & $22.33\pm0.58$ & SDSS DR12 & \cite{alam2021completed} \\
    0.51 & $D_M/r_d$ & $13.36\pm0.21$ & SDSS DR12 & \cite{alam2021completed} \\
    0.70 & $D_H/r_d$ & $19.33\pm0.53$ & eBOSS LRG & \cite{alam2021completed} \\
    0.70 & $D_M/r_d$ & $17.86\pm0.33$ & eBOSS LRG & \cite{alam2021completed} \\ 
    0.835 & $D_M/r_d$ & $18.92\pm0.51$ & DES Y3 & \cite{abbott2022desy3} \\
    0.85 & $D_V/r_d$ & $18.33^{+0.57}_{-0.62}$ & eBOSS ELG & \cite{alam2021completed} \\
    1.48 & $D_H/r_d$ & $13.26\pm0.55$ & eBOSS Quasar & \cite{alam2021completed} \\
    1.48 & $D_M/r_d$ & $30.69\pm0.80$ & eBOSS Quasar & \cite{alam2021completed} \\
    2.33 & $D_H/r_d$ & $8.93\pm0.28$ & eBOSS Ly$\alpha \times$ Ly$\alpha$ & \cite{bouroux2020lya} \\
    2.33 & $D_M/r_d$ & $37.6\pm1.9$ & eBOSS Ly$\alpha \times$ Ly$\alpha$ & \cite{bouroux2020lya} \\
\end{tabular}
\caption{\label{tab:nonDESI}A list of nonDESI BAO datasets used in this cosmological analysis. }
\end{ruledtabular}
\end{table*}

The sound horizon at the drag epoch, $r_d \equiv r_s(z_d)$, represents the true observable in BAO measurements. This feature appears in both the line-of-sight and transverse directions
\begin{eqnarray}
    \frac{D_H(z)}{r_d} &=& \frac{c}{H(z)r_d},\label{BAO_los} \\
    \frac{D_M(z)}{r_d} &=& \frac{r(z)}{r_d}, \label{BAO_trans}
\end{eqnarray}
where $r(z)$ is the same as transverse comoving distance $D_M(z)$\cite{hogg1999distance}, which depends on the spatial curvature to an object at redshift $z$ and is equal to $\chi(z)$ 
when the Universe is flat. 
The BAO measurements were also historically summarized by a single quantity representing the spherically averaged distance
\begin{equation} \label{BAO_combine}
    D_V(z) \equiv [zD_M^2(z)D_H(z)]^{1/3}.
\end{equation}

All BAO measurements used in this work can be divided into two categories. The first category, referred to as 'DESI BAO', comprises data from the DESI DR1 dataset \cite{adame2024desibao}, which includes observations from the bright galaxy sample (BGS), luminous red galaxies (LRG), emission line galaxies (ELG), quasars (QSO), and Lyman-$\alpha$ forest tracers, as shown in TABLE \ref{tab:DESI}. The second category, labeled 'nonDESI', encompasses BAO data from non-DESI surveys, mainly consisting of the complete Sloan Digital Sky Survey (SDSS) data release, which are summarized in TABLE I of Ref.\cite{perez2024nonDESIdata}, and this table is also presented in TABLE \ref{tab:nonDESI} of this paper.

\begin{figure*}[htb]
\centering
\includegraphics[width=0.90\textwidth,height=0.40\textheight]{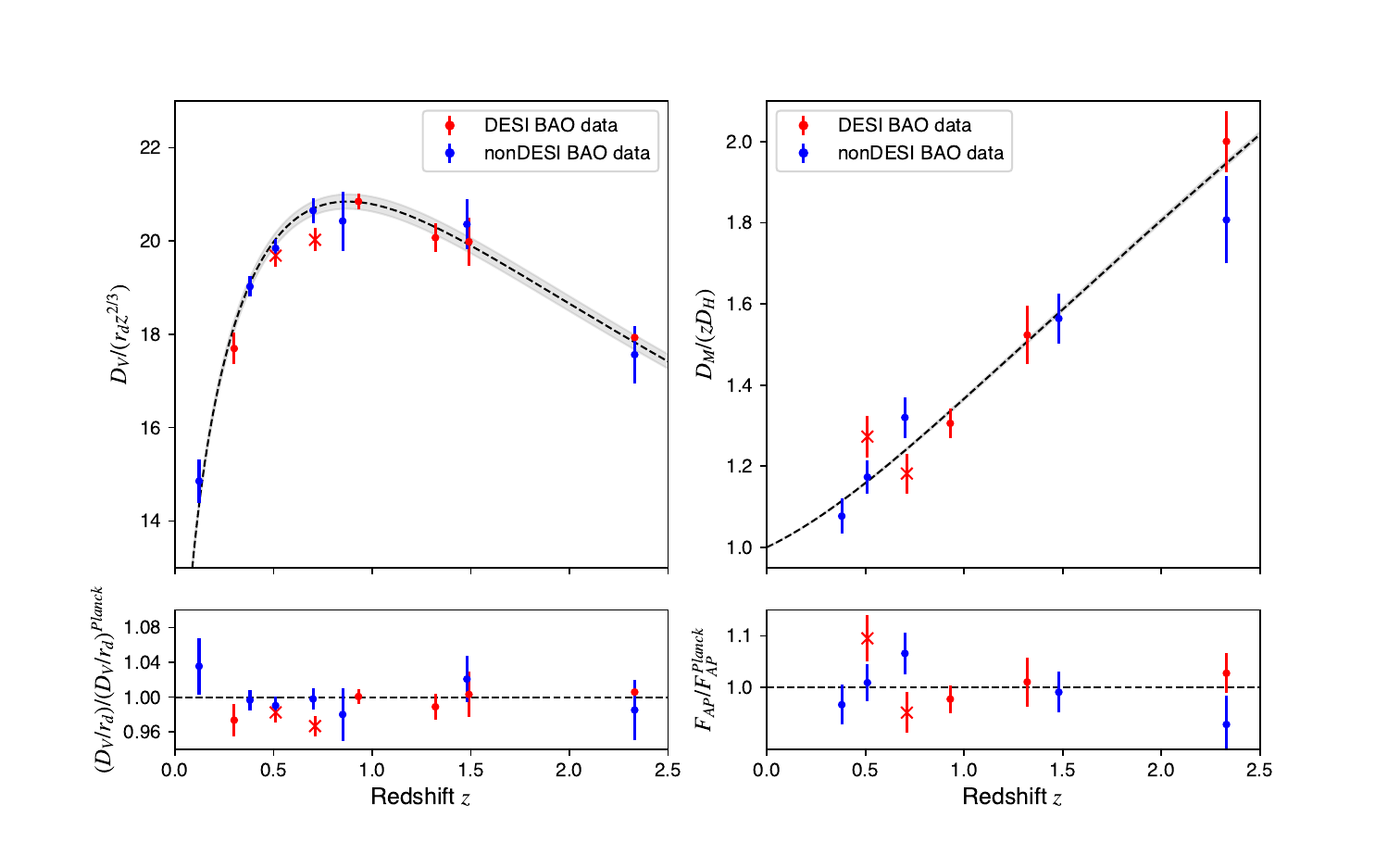}
\captionsetup{justification=raggedright}
\caption{The BAO measurements $D_V/r_d$ and $F_{\rm AP}=D_M/D_H$ at different redshifts are plotted in the top row's left and right panels, respectively. To enhance visual clarity and reduce the plot's dynamic range, a scaling of $z^{-2/3}$ has been applied to the left panel and $z^{-1}$ to the right panel. The dashed lines and grey bands represent model predictions from a $\Lambda$CDM model with parameters matching the Planck best-fit cosmology and the corresponding 68\% credible interval, respectively. In the DESI data points, the 'X' marker indicates that the tracer is LRG. The same data points, normalized by the predictions of the Planck best-fit flat $\Lambda$CDM model, are shown in the bottom row.} \label{Alldata}
\end{figure*}

These two datasets are depicted in FIG.\ref{Alldata}, where it can be observed that the DESI measurements for LRG1 ($z_{\rm eff}=0.51$) and LRG2 ($z_{\rm eff}=0.71$) show significant discrepancies with the BOSS DR12 results at similar redshifts, with differences of $1.6\sigma$ and $2.2\sigma$, respectively. The reasons for these inconsistencies are unknown \cite{adame2024desibao} and may be attributed to statistical fluctuation\cite{efstathiou2024challenges}. Therefore, we should be cautious in interpreting the conclusions regarding dynamical dark energy derived from the DESI data. If the evolution of dark energy is primarily determined by the measurements at these two redshifts, the dynamical dark energy signal might not indicate new physics but could instead be due to statistical reasons, such as sample variance fluctuation mentioned in Ref.\cite{adame2024desibao}. In the subsequent sections of this paper, we will analyze the impact of the measurements at these two redshifts on the inferred evolution of dark energy to distinguish whether the dynamical dark energy features in the DESI data are physical.

The likelihood of BAO, $\mathcal{L}_{\text{BAO}}$, takes different forms for different datasets. For most BAO data, including DESI datasets, a Gaussian likelihood function is used. However, for the SDSS Ly$\alpha$ forest data \cite{bouroux2020lya} and ELG analysis \cite{de2021elg}, the likelihood functions are non-Gaussian. Therefore, their full likelihoods need to be interpolated, which are available from the public SDSS repository\footnote{\url{https://svn.sdss.org/public/data/eboss/DR16cosmo/tags/v1_0_0/likelihoods/.}}.

\subsubsection{CMB}
CMB reflects the distance to the surface of last scattering $r(z_*)$ through the measured locations and amplitudes of the acoustic oscillation peaks. It measures two distance ratios: (i) $r(z_*)/r_s(z_*)$, the comoving distance to the last scattering surface divided by the sound horizon size at recombination, and (ii) $r(z_*)H(z_*)/c(1+z_*)$, the comoving distance divided by the comoving Hubble horizon size. These ratios are defined as the CMB shift parameters\cite{wang2007distanceprior-1,komatsu2009wmap1} $l_a$ and $R$, respectively. The \textit{Planck} Markov Chain Monte Carlo (MCMC) samples \cite{2020planck} can be compressed into three parameters, known as the CMB distance prior\cite{komatsu2009wmap1,wang2007distanceprior-1,wang2013distanceprior-2}, which includes $l_a$, $R$, and the physical baryon density $\omega_b$. This allows the use of the CMB distance prior to replace the full CMB power spectrum for simplicity in constraining cosmological parameters
\begin{eqnarray}
    R&=&\sqrt{\Omega_mH^2_0}r(z_*)/c, \label{CMB_prior1} \\
    l_a&=&\pi r(z_*)/r_s(z_*). \label{CMB_prior2}
\end{eqnarray}
The parameter $z_*$ can be calculated by the fitting formula \cite{hu1996analyticalformula}
\begin{equation}\label{z*}
    z_*=1048[1+0.00124\omega^{-0.738}_b][1+g_1\omega^{g_2}_m],
\end{equation}
where
\begin{equation}\label{z*1}
    g_1=\frac{0.0783\omega^{-0.238}_b}{1+39.5\omega^{0.763}_b},~~ g_2=\frac{0.560}{1+21.1\omega^{1.81}_b}.
\end{equation}

The final result, which is extracted from \textit{Planck} 2018 TT,TE,EE+lowE chains, can be transformed in terms of a data vector $\bm{v}=(R,l_a,\omega_b)^T$ \footnote{Actually, spectral index of the primordial power spectrum $n_s$ is entailed for completeness, however we neglect $n_s$ because it doesn't relate to our constraint.} and their covariance matrix. For a flat universe \cite{wang2013distanceprior-2}, this data vector is
\begin{equation}\label{CMB_data}
\bm{v}=\begin{pmatrix}
    1.74963 \\
    301.80845 \\
    0.02237 \\
    \end{pmatrix}
\end{equation}
and the covariance matrix is
\begin{equation}\label{CMB_cov}
    \bm{C_v}=10^{-8}\times 
    \begin{pmatrix}    
        1598.9554 & 17112.007 & -36.311179  \\
        17112.007 & 811208.45 & -494.79813  \\
       -36.311179 & -494.79813& 2.1242182 
    \end{pmatrix} .\nonumber
\end{equation}

The loglikelihood function for CMB can be written in the form:
\begin{equation}\label{L_CMB}
\mathcal{L}_{CMB} = -\frac{1}{2}\chi^2_{CMB} = -\frac{1}{2}(\bm{v}-\bm{v}_{th})^T \bm{C_v}^{-1} (\bm{v}-\bm{v}_{th}).
\end{equation}
It is necessary to note that the CMB distance priors extracted from the full CMB data are independent of the model of dark energy assumed \cite{wang2019distanceprior-3}. This implies that the CMB distance priors Eq.\eqref{CMB_data} and Eq.\eqref{CMB_cov}, although derived from flat $\Lambda$CDM chains, can be applied to the analysis of any dark energy model.

\subsubsection{SN Ia}
Type Ia supernova (SN Ia) are characterized by an almost uniform intrinsic luminosity, making them reliable standard candles. We compare the theoretical distance moduli $\mu_{\rm th}$ with the observed distance moduli $\mu_{\rm obs}$ of SN Ia from the Pantheon$+$ sample \cite{brout2022pantheon+}. The theoretical distance moduli are defined using the luminosity distance $d_L(z) = (1+z)r(z)$
\begin{equation}\label{mu_th}
\mu_{\rm th}=5\log_{10}\left(\frac{d_L(z)}{\rm Mpc}\right)+25,
\end{equation}
The observed distance modulus, $\mu_{\rm obs} = m_{\rm obs} - M$, includes the observed apparent magnitude of an SN Ia, $m_{\rm obs}$, and the fiducial magnitude, $M$. The fiducial magnitude can be calibrated by setting an absolute distance scale using primary distance anchors. In the Pantheon$+$ samples, the SH0ES Cepheid host distance anchors \cite{SH0ES2022} are included in the likelihood, effectively incorporating a $H_0$ prior measured by SH0ES collaboration.

Using the formalism developed in Ref.\cite{conley2010supernova}, we can constrain cosmological parameters with a compilation of 1550 supernovae from the Pantheon$+$ sample, spanning redshifts from 0.01 to 2.26
\begin{equation}\label{L_SNIa}
\mathcal{L}_{SN} = -\frac{1}{2}\chi^2_{SN} = -\frac{1}{2}(\bm{\mu_{\rm obs}}-\bm{\mu_{\rm th}})^T \bm{C}^{-1} (\bm{\mu_{\rm obs}}-\bm{\mu_{\rm th}}),
\end{equation}
where $C$ is the SN covariance matrix, which combines statistical and systematic contribution.

\begin{table*}[htb]
\captionsetup{justification=raggedright}
\begin{ruledtabular}
\begin{tabular}{cc|cc}
 Datasets & ${\rm ratio}(\omega_m)$ & Datasets & ${\rm ratio}(\omega_m)$ \\ \hline
    DESI BAO & $1.0171\pm0.0066(2.6\sigma)$ & nonDESI BAO & $1.0082\pm0.0068(1.2\sigma)$ \\
    DESI BAO (w/o LRG1) & $1.0174\pm0.0074(2.4\sigma)$ & DESI BAO (w/o LRG2) & $1.0124\pm0.0071(1.7\sigma)$ \\
    DESI BAO (w/o LRG1,2) & $1.0100\pm0.0082(1.2\sigma)$ & DESI BAO (w/o LRG2\ $D_V$) & $1.0129\pm0.0071(1.8\sigma)$ \\
\end{tabular}
\caption{\label{tab:ratio} The posterior constraint of ${\rm ratio}(\omega_m)$ with $1\sigma$ (68\%) credible-interval are shown. In the "Datasets" column, "CMB" is omitted to highlight the differences in BAO data. In the "${\rm ratio}(\omega_m)$" column, values in parentheses indicate the discrepancy from the predicted value of 1.}
\end{ruledtabular}
\end{table*}

\subsection{Analysis} 
We perform a MCMC sampling of the parameter space using the \texttt{emcee} sampler \cite{emcee} to constrain the parameters. For the three analysis methods introduced above, we will discuss the parameter and prior choices in the following parts
\begin{itemize}
    \item \textbf{${\rm ratio}(\omega_m)$ analysis:}\ The complete parameter set for this constraint is $\lbrace \tilde{\omega}_b, \tilde{\omega}_m, \Omega_m, {\rm ratio}(\omega_m) \rbrace$, using observational data from BAO and CMB. The corresponding priors are $\tilde{\omega}_b \in (0.001, 0.3)$, $\tilde{\omega}_m \in (0.05, 0.5)$, $\Omega_m \in (0.1, 0.9)$, and ${\rm ratio}(\omega_m) \in (0.1, 10)$.
    \item \textbf{Taylor Reconstruction of $H(z)$:}\ This parameter constraint involves three different data combinations, denoted as `BAO', `BAO+SNIa', and `BAO+H0+SNIa'. Here, `H0' refers to the SH0ES\cite{SH0ES2022} direct measurement of $H_0$ as $73.04 \pm 1.04\ \rm km\ s^{-1}\ Mpc^{-1}$. If only BAO data are used, the parameter set is $\lbrace \alpha_0, \alpha_1, \alpha_2, \alpha_3, \alpha_4 \rbrace$. Combining SN Ia data introduces a new parameter, $M$. Thus, for 'BAO+SNIa', the parameter set is $\lbrace \alpha_0, \alpha_1, \alpha_2, \alpha_3, \alpha_4, M \rbrace$. Including $H_0$ information breaks the degeneracy between $H_0$, $r_d$, and $M$. Therefore, for `BAO+H0+SNIa', the parameter set is $\lbrace \alpha_0, \alpha_1, \alpha_2, \alpha_3, \alpha_4, r_d, M \rbrace$. Note that no priors are assumed in this parameter estimation.
    \item \textbf{Chebyshev Reconstruction of $w(z)$:}\ This parameter constraint involves three different data combinations, denoted as 'BAO', 'BAO+SNIa', and 'BAO+SNIa+CMB'. Since CMB provides pre-recombination information, $r_d$ is no longer a free parameter. Thus, the parameter set is $\lbrace C_0, C_1, C_2, C_3, \Omega_m, H_0r_d \rbrace$ when considering non-CMB observations, and $\lbrace C_0, C_1, C_2, C_3, \Omega_m, \Omega_b, H_0 \rbrace$ when including CMB likelihoods. The absolute magnitude $M$ is treated as a nuisance parameter when incorporating Pantheon$+$ samples. The priors used in this analysis are the same as those in Ref.\cite{calderon2024desi}: Gaussian priors for $C_i \sim \mathcal{N}(\mu^{\Lambda \text{CDM}}, \sigma^2)$, where $\mu^{\Lambda \text{CDM}}$ corresponds to $C_0 = 1$ and $C_{i>0} = 0$; and uniform priors for other physical parameters $\Omega_m \in (0.01, 0.99)$, $H_0 \in (20, 100)$, and $H_0r_d \in (3650, 18250)$.
\end{itemize}

\section{Results and Discussion} \label{sec3}

\begin{figure}[htb]
\centering
\includegraphics[scale=0.45]{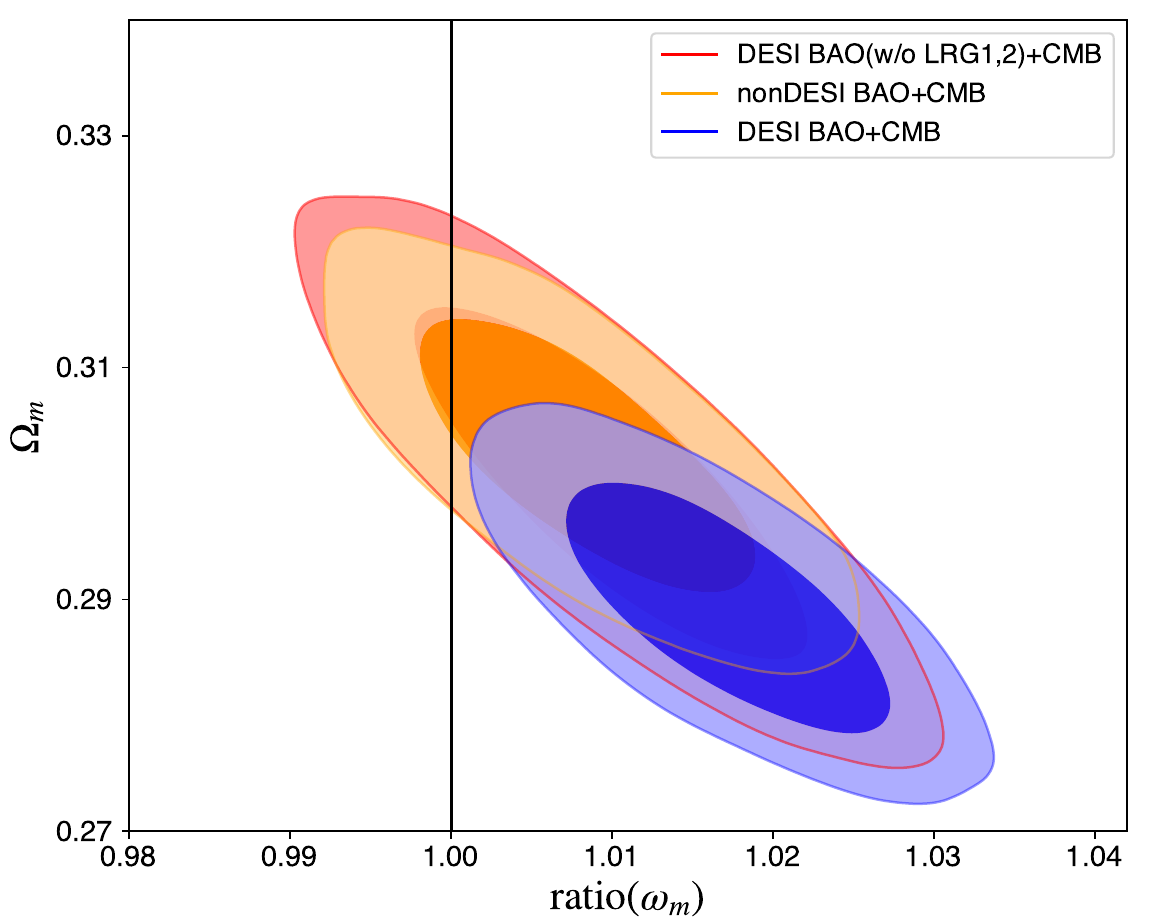}
\captionsetup{justification=raggedright}
\caption{68\% and 95\% credible-interval contours for parameters ${\rm ratio}(\omega_m)$ and $\Omega_m$ obtained under a flat $\Lambda$CDM framework from fits to the combination of different BAO measurements and CMB data. "w/o" means this portion of BAO data was excluded from the parameter estimation.}\label{ratio_plot}
\end{figure}

In this section, we present our findings using different approaches and data combinations.

\subsection{${\rm ratio}(\omega_m)$ Analysis}
The constrained contours of the parameters using BAO+CMB are illustrated in FIG.\ref{ratio_plot}, and the results are also summarized in TABLE \ref{tab:ratio}. As mentioned above, since $\omega_m$ is time-independent within the $\Lambda$CDM framework, the value of an early and late cosmological parameter should be identical, implying that the predicted value of ${\rm ratio}(\omega_m)$ is 1. However, the constraint obtained from DESI BAO+CMB is ${\rm ratio}(\omega_m) = 1.0171 \pm 0.0066$, indicating a $2.6\sigma$ tension with $\Lambda$CDM. For comparison, the corresponding result from nonDESI data is ${\rm ratio}(\omega_m) = 1.0082 \pm 0.0068$, and the difference between this result and the DESI result is clearly illustrated in FIG. \ref{ratio_plot}. According to the DESI collaboration \cite{adame2024desibao}, DESI BAO data favor solutions with $w_0 > -1$ and $w_a < 0$. The level of tension with the $\Lambda$CDM model can reach up to $3.9\sigma$ when combining DESI and CMB information with the DESY5 SN Ia dataset. These two consistent conclusions demonstrate that ${\rm ratio}(\omega_m)$ can accurately reflect the deviation from the standard cosmological model.

It is noted that the DESI LRG1 ($z_{\rm eff}=0.51$) and LRG2 ($z_{\rm eff}=0.71$) data are in tension with the results from BOSS DR12 at similar redshifts, showing discrepancies of $1.6\sigma$ and $2.2\sigma$, respectively. These data points are crucial for determining the evolution of dark energy, as they are near the dark energy-dominated epoch (around $z \approx 0.5$). We believe that the combination of DESI LRG1 and LRG2 data leads to the observed deviation from $w = -1$, indicating a departure from the $\Lambda$CDM model. When we exclude this combination of DESI data, we find ${\rm ratio}(\omega_m) = 1.0100 \pm 0.0082$, reducing the tension from $2.6\sigma$ to $1.2\sigma$, suggesting a preference for $\Lambda$CDM and supporting our hypothesis.

However, the authors of Ref.\cite{wang2024role} found that the angle-averaged distance $D_V$ in LRG2 predominantly favors the presence of dynamical dark energy through a step-by-step exclusion of data points. Excluding the measurement $D_V$ of LRG2, we find ${\rm ratio}(\omega_m) = 1.0129 \pm 0.0071$, indicating that the predicted value of $\Lambda$CDM ${\rm ratio}(\omega_m) = 1$ is still outside the 93\% credible interval. Similar results were obtained for LRG1. We attribute this to the fact that the analysis in Ref.\cite{wang2024role} was conducted under a $w_0w_a$ cosmology. Therefore, we believe it is necessary to use model-independent methods to reconstruct $H(z)$ and provide a robust constraint on the evolution of $w(z)$.


\subsection{Model-independent Reconstruction Analysis}

\begin{figure}[htb]
\centering
\includegraphics[scale=0.45]{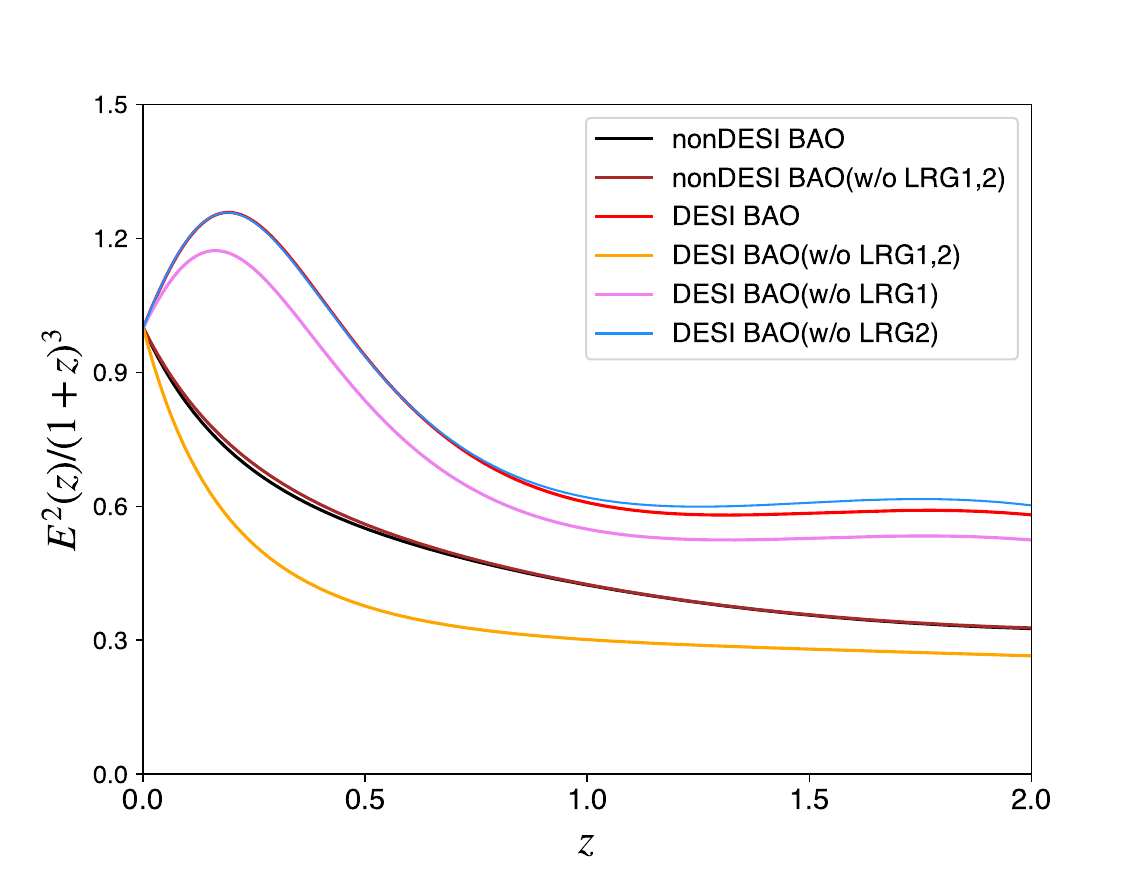}
\captionsetup{justification=raggedright}
\caption{Taylor reconstruction of $E(z)$ using different BAO datasets, divided by $(1+z)^3$ to highlight the time evolution characteristics of the dark energy component.} \label{TR_E}
\end{figure}

\subsubsection{$E(z)$ Reconstruction}


\begin{figure*}[htb]
\centering
\includegraphics[width=\linewidth]{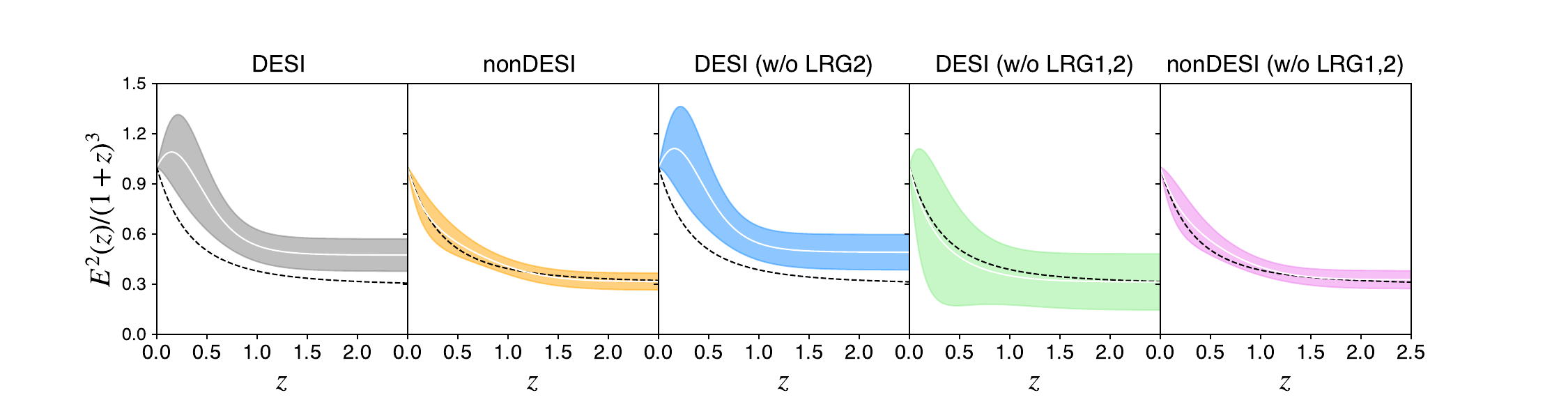}
\captionsetup{justification=raggedright}
\caption{Chebyshev reconstruction of $E(z)$ using different BAO datasets, divided by $(1+z)^3$ for comparison with TR results in FIG. \ref{TR_E}. White lines show the medians of the posterior distributions, shaded regions indicate the 68\% credible intervals, and black-dashed lines represent the best-fit $\Lambda$CDM predictions.} \label{CR_E}
\end{figure*}

When using only BAO data for Taylor reconstruction of $H(z)$, the measurements are $D_M(z)/r_d$ and $H(z)r_d$, lacking information about $r_d$. Consequently, the constrained parameter $\alpha_0$ is actually $\alpha_0/r_d$ in physical terms, meaning the reconstruction yields $H(z)r_d$ instead of $H(z)$. Since we are interested in the evolution feature of dark energy reflected by the observational data, we reconstruct $E(z)$ using the obtained $\alpha_1 - \alpha_4$, and divide by $(1+z)^3$ to exclude the matter component's influence. The reconstruction results derived from different BAO datasets are shown in FIG.\ref{TR_E}. In the $\Lambda$CDM framework (including $w$CDM model), $E^2(z)/(1+z)^3 \propto f_{\rm DE}(z)/(1+z)^3 = (1+z)^{3w}$ should decrease with increasing $z$, as shown by the curve corresponding to nonDESI BAO data. 

However, the curve for DESI BAO behaves anomalously, increasing with $z$ in the redshift range $0-0.5$ and decreasing in subsequent ranges, indicating a transition from $w(z) > 0$ to $w(z) < 0$. This curve of DESI BAO clearly exhibits the dynamical features of dark energy. We find that the evolution of the DESI curve remains unchanged whether LRG1 or LRG2 is individually removed, consistently showing features of non-cosmological constant dark energy. However, when both LRG1 and LRG2 are removed, the DESI curve aligns with the nonDESI curve, indicating $w(z)$ is negative. To verify these conclusions, we performed Chebyshev reconstruction using the same data, with results shown in FIG.\ref{CR_E}. The results from both model-independent methods are essentially the same. The black dashed line, representing the $\Lambda$CDM prediction, and the white curve, representing the reconstructed evolution of $E(z)$, nearly overlap when using nonDESI BAO and DESI BAO data with LRG1 and LRG2 removed.  This confirms our conclusion from the ${\rm ratio}(\omega_m)$ analysis: the combination of LRG1 and LRG2 in the DESI 2024 BAO data leads to the departure from $w = -1$, i.e., the departure from $\Lambda$CDM.

\subsubsection{Other Functions Reconstruction}


\begin{figure*}[htb]
\centering
\includegraphics[width=\linewidth, height=0.48\textheight]{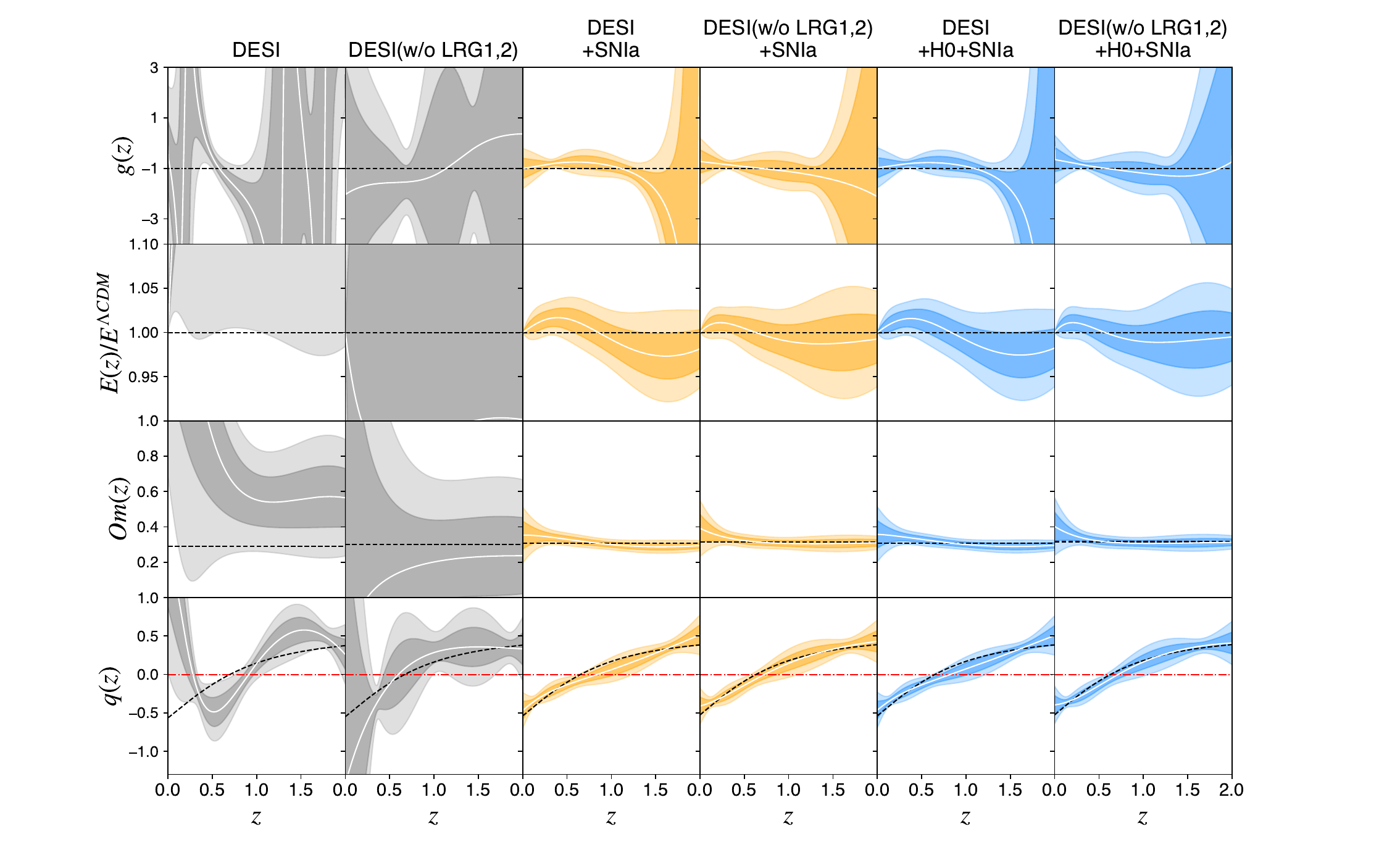}
\captionsetup{justification=raggedright}
\caption{The evolution of four terms derived from Taylor reconstructions of $H(z)$ using different datasets, including Pantheon$+$ and the $H_0$ measurement by SH0ES, to determine the impact of LRG1 and LRG2 in DESI BAO data. White lines show the medians of the posterior distributions, shaded regions indicate the 68\% and 95\% credible intervals, black-dashed lines represent the best-fit $\Lambda$CDM predictions, and red dashed lines in the $q(z)$ panels indicate $q = 0$ to distinguish between an accelerating and decelerating universe.} \label{TR_all}
\end{figure*}

\begin{figure*}[htb]
\centering
\includegraphics[width=\linewidth, height=0.50\textheight]{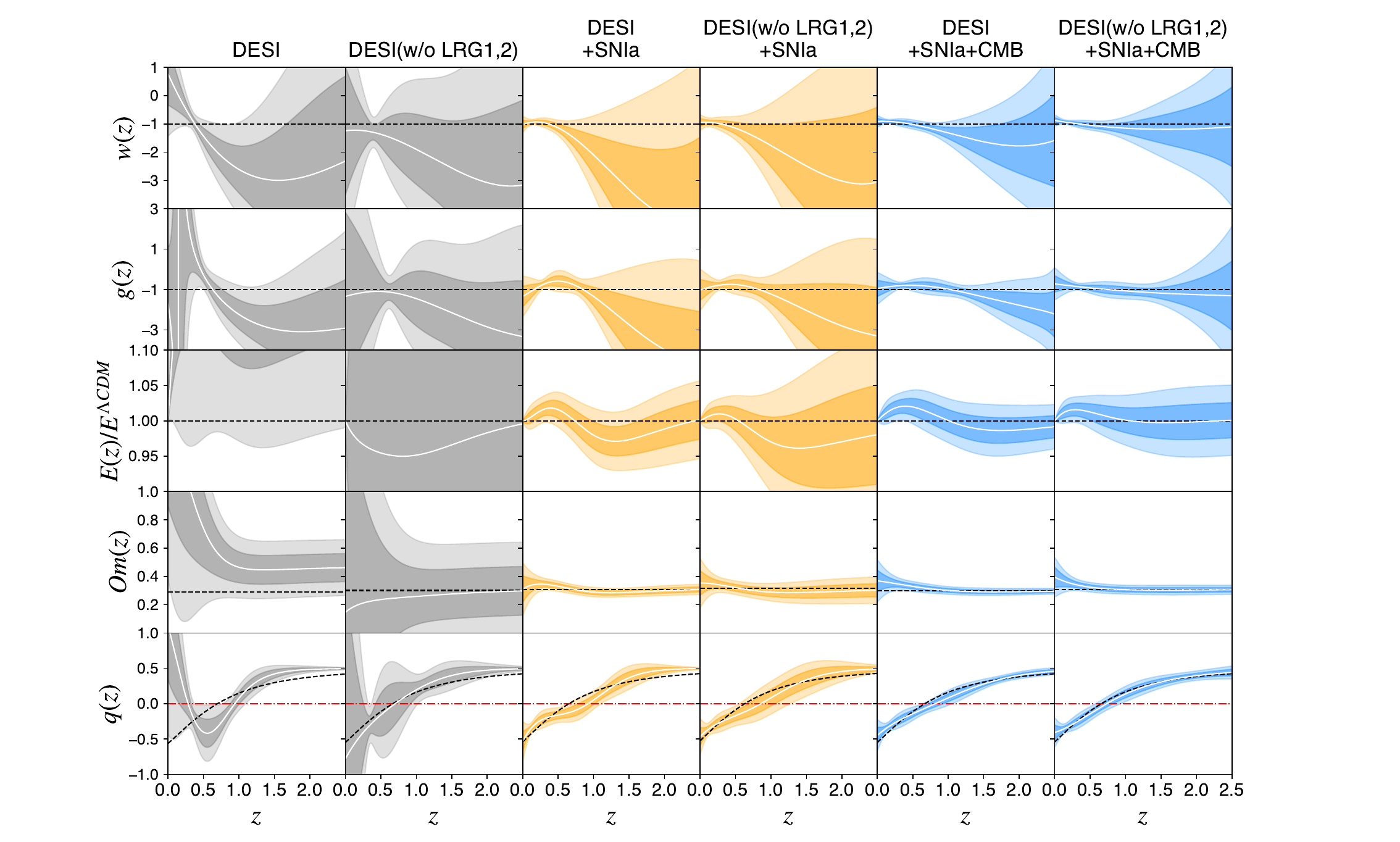}
\captionsetup{justification=raggedright}
\caption{The evolution of five terms derived from Chebyshev reconstructions of $w(z)$ using different datasets, including Pantheon$+$ and \textit{Planck} CMB data, to determine the impact of LRG1 and LRG2 in DESI BAO data. White lines show the medians of the posterior distributions, shaded regions indicate the 68\% and 95\% credible intervals, black-dashed lines represent the best-fit $\Lambda$CDM predictions, and red dashed lines in the $q(z)$ panels indicate $q = 0$ to distinguish between an accelerating and decelerating universe.} \label{CR_all}
\end{figure*}

We introduce four additional physical quantities: the characterization function of dark energy $g(z)$, the relative $E(z)$ normalized by $E^{\Lambda \text{CDM}}$, the diagnostic $Om(z)$, and the deceleration function $q(z)$. These functions can be expressed in terms of $H(z)$ obtained from Taylor or Chebyshev reconstructions. By incorporating SN Ia and CMB data alongside BAO, the shapes of these functions allow us to analyze potential features of dynamical dark energy from a richer dataset.

Following Ref.\cite{wang2024dynamical}, we introduce the characterisation function of dark energy, $g(z)$ \cite{sahni2003g(z)_1, alam2003g(z)_2}
\begin{equation} \label{g}
    g(z) \equiv \frac{1}{3}\left[\frac{(1+z)f{''}(z)}{f{'}(z)}+1\right]= w+\frac{1+z}{3}\frac{w{'}}{w},
\end{equation}
where $f(z) \equiv H^2(z)/(1+z)^3$ and the prime symbol ($'$) denotes the first derivative of the function. The value of $g(z)$ can be used as a diagnostic for dark energy models. For example, $\Lambda$CDM requires $g(z)$ to be identically $-1$. 

Following Ref.\cite{calderon2024desi}, we also use $E(z)/E^{\Lambda \text{CDM}}$ to show the constraints on the evolution of the expansion history $E(z)$, normalized to the best fit value of $\Lambda$CDM for visual clarity.

The $Om(z)$ diagnostic, developed in Ref.\cite{sahni2008Om}, is used to distinguish dark energy models and is defined as
\begin{equation} \label{Om}
    Om(z) \equiv \frac{E^2(z)-1}{(1+z)^3-1}.
\end{equation}
It will revert to $\Omega_m$ under the assumption of $\Lambda$CDM. 

The final one is the deceleration function, which describes whether the expansion of the universe is accelerating or decelerating
\begin{equation} \label{q}
    q \equiv -\frac{\Ddot{a}\  a}{\dot{a}^2}= \frac{1}{2}\left[1+\frac{(1+z)f^{'}(z)}{f(z)}\right],
\end{equation}
where $a$ is the scale factor, and its relation to the redshift $z$ is $a = 1/(1+z)$.

The results corresponding to the two model-independent approaches are presented in FIG.\ref{TR_all} and FIG.\ref{CR_all}. In each panel, the medians of the posterior distributions are shown as white lines, with shaded areas indicating the 68\% and 95\% credible intervals. The best-fit $\Lambda$CDM predictions are depicted by black-dashed lines, while red dashed lines in the $q(z)$ panels highlight $q = 0$ to distinguish between an accelerating and decelerating universe.

The Taylor reconstruction results, using DESI BAO data, Pantheon$+$ SN Ia data, and SH0ES $H_0$ measurement, are shown in FIG.\ref{TR_all}. The $g(z)$ derived from the full DESI BAO data exhibits significant variability due to the presence of extrema in the corresponding $f(z)$ (as shown in FIG.\ref{TR_E}), where $f{'}(z)$ equals zero in the denominator, rendering $g(z)$ unable to provide meaningful information about the evolution of $w(z)$. However, when LRG1 and LRG2 are removed from the DESI data, the singularities in $g(z)$ disappear, and the results align with the $\Lambda$CDM predictions within a large margin of error. Similarly, the DESI BAO data in the redshift range $0-0.5$ shows more than a $2\sigma$ deviation from $\Lambda$CDM in $E(z)/E^{\Lambda \text{CDM}}$, $Om(z)$, and $q(z)$. This deviation disappears when this combination is excluded. Including SN Ia data significantly reduces the uncertainties in all functions and shows consistent values with $\Lambda$CDM predictions. Adding the $H_0$ prior does not cause significant changes in the evolution of any physical quantities.

The Chebyshev reconstruction results, which replace the SH0ES $H_0$ prior with \textit{Planck} CMB data, are shown in FIG.\ref{CR_all}.  The shapes of $E(z)/E^{\Lambda \text{CDM}}$, $Om(z)$, and $q(z)$ obtained from the Chebyshev reconstruction of $w(z)$ are essentially consistent with the Taylor reconstruction results, thus we are more interested in the evolution of $w(z)$ and $g(z)$. DESI BAO data prefer a transition from quintessence ($w > -1$) to phantom ($w < -1$) dark energy, referred to as quintom\cite{feng2005quintom1,guo2005quintom2l} dark energy. After including SN Ia and CMB data, the dynamic features of $w(z)$ become weaker, tending towards a phantom cosmology. When LRG1 and LRG2 are removed from DESI, the phantom dark energy phenomenon persists. However, we find that the shape of $w(z)$ is almost identical to the corresponding $g(z)$, indicating $w' = 0$, which suggests non-dynamic feature of dark energy. When CMB data is combined, the precision of the physical quantities significantly improves, and both $w(z)$ and $g(z)$ become nearly identical to the $\Lambda$CDM predictions. Through the reconstruction of these physical functions using SN Ia and CMB data, the importance of LRG1 and LRG2 in the evolution of dark energy is reaffirmed.


\section{Conclusions} \label{sec4}  
DESI DR1 BAO observations, with their extensive sample size, provide an unprecedented opportunity to investigate the universe's background and gain deeper insights into the feature of dark energy. According to the DESI collaboration, there is a significant preference for dynamical dark energy, observed at confidence levels of $2.5\sigma$, $3.5\sigma$, and $3.9\sigma$ when DESI data is combined with Planck CMB, Pantheon$+$, Union3, and DESY5 SN samples, respectively. We should approach the conclusions of evolving dark energy derived from DESI data with caution, as the reasons for the inconsistencies between the DESI LRG1 ($z_{\rm eff}=0.51$) and LRG2 ($z_{\rm eff}=0.71$) data and the BOSS DR12 results at similar redshifts are unknown. Considering that these redshifts fall within the dark energy-dominated epoch, if the combination of DESI LRG1 and LRG2 data significantly contributes to the observed dynamical features of dark energy, then the dynamic dark energy observed in DESI data might be of non-physical origin. Therefore, we need to explore the impact of this combination on the evolution of dark energy.

The deviation observed in the cosmological constant $\Lambda$ indicates a departure from the standard cosmological model, which can be quantified by the parameter ${\rm ratio}(\omega_m)$. We find ${\rm ratio}(\omega_m) = 1.0171 \pm 0.0066$ for DESI and ${\rm ratio}(\omega_m) = 1.0100 \pm 0.0082$ for DESI excluding LRG1 and LRG2, showing a reduction from $2.6\sigma$ to $1.2\sigma$, supporting our hypothesis. Subsequently, we calculated ${\rm ratio}(\omega_m)$ for DESI with each set of observations removed individually and found no significant change compared to the value obtained from the full DESI data, ruling out the possibility that LRG1 or LRG2 alone causes $w \neq -1$. To validate this conclusion, we used two model-independent methods to extract information on dark energy evolution from the observational data. Our reconstructions reveal that this data combination dictates the evolution of dark energy and the late-time expansion history of the universe: if the DESI LRG1 and LRG2 measurements are accurate, quintom cosmology is preferred; if the BOSS measurements at these redshifts are accurate, the $\Lambda$CDM model remains consistent with observations without modification. Therefore, future precise measurements of the BAO features in these redshift ranges will determine the destiny of the $\Lambda$CDM model. 

~

\begin{acknowledgments}
This work is supported by the National Key R\&D Program of China (Grant No. 2021YFC2203102 and 2022YFC2204602), Strategic Priority Research Program of the Chinese Academy of Science (Grant No. XDB0550300), the National Natural Science Foundation of China (Grant No. 12325301 and 12273035), the Fundamental Research Funds for the Central Universities (Grant No. WK2030000036 and WK3440000004), the Science Research Grants from the China Manned Space Project (Grant No.CMS-CSST-2021-B01), the 111 Project for "Observational and Theoretical Research on Dark Matter and Dark Energy" (Grant No. B23042).
\end{acknowledgments}


\bibliography{reference}

\end{document}